\documentclass[
reprint,
showpacs,preprintnumbers,
nofootinbib,
 amsmath,amssymb,
 aps,
]{revtex4-1}
\usepackage{multirow}
\usepackage{enumerate}
\usepackage{color}
\usepackage{graphicx}% Include figure files
\usepackage{dcolumn}% Align table columns on decimal point
\usepackage{bm}% bold math
\usepackage{amssymb}
\usepackage{subfigure}
\begin{document}
\title{Massive charged Dirac fields around Reissner-Nordstr\"{o}m black holes: quasibound states and long-lived modes}
\author{Yang Huang}
\email{saisehuang@163.com}
\author{Dao-Jun Liu}%
 \email{djliu@shnu.edu.cn}
  \author{Xiang-hua Zhai}%
  \email{zhaixh@shnu.edu.cn}
  \author{Xin-zhou Li}%
  \email{kychz@shnu.edu.cn}
  \affiliation{Center for Astrophysics, Shanghai Normal University, 100 Guilin Road, Shanghai 200234, China
%\affiliation{%
}%
\affiliation{Department of Physics, Shanghai Normal University, 100 Guilin Road, Shanghai 200234, China}
\date{\today}% It is always \today, today,

\begin{abstract}
	The behavior of a massive charged test Dirac field in the background of a Reissner-Nordstr\"{o}m black hole is investigated. Especially, we obtain the frequencies of quasibound states by solving the Dirac equation numerically both in time and frequency domain. Our results suggest that although the absence of superradiance excludes the existence of stationary solutions for massive Dirac fields, it is still possible to find arbitrarily long-lived solutions.
   
%\begin{description}
%\end{description}
\end{abstract}

%\pacs{04.30.Nk, 04.70.Bw}% PACS, the Physics and Astronomy Classification Scheme.
%\keywords{Suggested keywords}%Use showkeys class option if keyword display desired
\maketitle

%\tableofcontents
\section{Introduction}
In classical general relativity, according to the no-hair conjecture \cite{doi:10.1063/1.3022513,PhysRevLett.28.452}, all asymptotically flat stationary black hole (BH) solutions can be completely characterized by only three parameters: mass, electric charge, and angular momentum. A perturbed BH will return to a stationary state and all other external fields will inevitably decay with time.

Although the no-hair conjecture forbids any nontrivial field distribution in BH spacetimes, it does not rule out the existence of dynamical solutions that decay very slowly\cite{PhysRevD.84.083008}. In Refs.\cite{PhysRevD.84.083008,PhysRevLett.109.081102}, Barranco \textit{et al.} studied the evolution of massive scalar fields around a Schwarzschild BH and found configurations that can survive for a very long time, even for cosmological time scales. Zhou \textit{et al.} further showed that such long-lived nontrivial distributions can be extended to Dirac fields \cite{PhysRevD.89.043006}. These dynamical resonance solutions are basically described by the so-called quasibound states of massive fields in the BH spacetimes, which have attracted much attention in recent years \cite{PhysRevD.88.023514,PhysRevD.85.044043,0264-9381-32-18-184001,Damour1976,ZOUROS1979139,Furuhashi:2004jk,PhysRevD.70.044039,PhysRevD.76.084001,PhysRevD.87.124026,PhysRevD.22.2323,PhysRevD.86.104017,PhysRevD.72.105014,doi:10.1142/S0218271817501413,0264-9381-34-15-155002}. Recent studies have shown that these resonances will dominate the dynamics of the massive fields in BH spacetimes, and could affect the late time waveforms of the gravitational radiation \cite{PhysRevD.89.104059,Barausse:2014pra,PhysRevD.90.065019}.

A common feature of the results in Refs.\cite{PhysRevD.84.083008,PhysRevLett.109.081102,PhysRevD.89.043006} is that the long-lived modes are obtained in the limit of small-mass coupling $M\mu\ll1$, where $M$ and $\mu$ are the masses of the BH and field, respectively. There is a simple physical explanation for this: smaller value of $M\mu$ corresponds to weaker gravitational attraction between the external field and BH, which implies that it takes a longer time for the BH to absorb the external fields around it. However, the small mass coupling requirement for long-lived field configurations is no longer necessary in at least two scenarios.

Firstly, massive bosonic fields in rotating BH spacetimes can form stationary clouds, which correspond to the intermediate states between the decaying and superradiantly growing ones \cite{Hod.2013zza,PhysRevD.86.104026,Hod2017,PhysRevD.94.064030,Delgado2016234,PhysRevLett.112.221101}. It is not possible to find such stationary clouds for Dirac fields in BH spacetimes due to the absence of superradiance \cite{PhysRevD.18.4799,PhysRevD.15.3060}. However, for Dirac fields around a rapidly rotating Kerr BH, the decay rate of the co-rotating mode is suppressed, so long as the real part of the bound state frequency obeys a 'superradiant'-like condition \cite{0264-9381-32-18-184001}.

Secondly, for a massive charged scalar field on the Reissner-Nordstr\"{o}m (RN) background, the decay rate of the field configuration could be extremely slow when $M\mu\gtrsim qQ$ \cite{PhysRevD.90.064004,Degollado2013}, where $q$ and $Q$ are the charges of the field and BH, respectively. Although scalar fields on the RN background are subject to amplification by a charged version of superradiance, no stationary scalar modes exist, because the superradiance and quasibound state conditions are incompatible with each other in this case \cite{PhysRevD.91.044047,Hod:2013eea,Hod:2013nn}.

The study of a Dirac test field in a black hole background is a subject of long-standing interest and the Dirac-Maxwell-(Einstein) system has been
considered in a series of paper by Finster et. al., see e.g.~\cite{Finster:1998ux,Finster:1998ak}. Among other things, these papers establish the absence of normalizable, time-periodic solutions in a RN background, together with the existence of solitonic solutions sustained by gravity effects.

In this paper, we shall consider a classical massive charged Dirac field propagating on the RN background and intend to work out whether long-lived Dirac configurations surrounding a RN BH exist. The organization of this paper is as follows. In Sec.\ref{Sec: background end equation of motion}, the separation of variables procedure for massive Dirac fields around RN BHs is reviewed. Then, in Sec.\ref{Sec: Frequency domain method} and Sec.\ref{Sec: Time domain method}, we compute the quasibound state solutions by solving the Dirac equation on the RN background in frequency and time domains, respectively. Finally, the paper ends up with a discussion and conclusion in Sec.\ref{Sec: conclusion}. Throughout the paper, we use natural units in which $G=c=\hbar=1$.

\section{Massive Dirac fields around Reissner-Nordstr\"{o}m black holes}\label{Sec: background end equation of motion}
\subsection{The Dirac equation in Reissner-Nordstr\"{o}m black hole spacetimes}
We start by the Dirac equation in RN BH spacetime. The RN metric, in Boyer-Lindquist coordinates, is given by
\begin{equation}\label{Eq: the line element}
	ds^2=-\frac{\Delta}{r^2}dt^2+\frac{r^2}{\Delta}dr^2+r^2\left(d\theta^2+\sin^2\theta d\phi^2\right),
\end{equation}
where $\Delta=r^2-2Mr+Q^2$. The electromagnetic potential of the charged black hole reads $A_\mu=(-Q/r,0,0,0)$. The RN BH has two horizons, $r_\pm=M\pm\sqrt{M^2-Q^2}$. When $Q=0$, the potential $A_\mu$ and the inner horizon vanish, and the BH reduces to a Schwarzschild one.

Massive charged Dirac field in curved spacetime is described by \cite{RevModPhys.29.465}
\begin{equation}\label{Eq: The Dirac eq in curved spacetime}
	\left(\gamma^{\mu}D_\mu-\mu\right)\bm{\Psi}=0,
\end{equation}
where $\mu$ is the mass of the field, $\bm{\Psi}$ is the Dirac four-spinor and $\gamma^\mu$ are coordinate-dependent Dirac four-matrices, whose components are the functions of spacetime coordinates. The spinor covariant derivatives in Eq.(\ref{Eq: The Dirac eq in curved spacetime}) are given by
\begin{equation}
	D_\mu=\partial_\mu-\Gamma_\mu-iqA_\mu,
\end{equation}
where $q$ is the charge of the Dirac field, and $\Gamma_\nu$ are the spinor connection matrices. The separation of variables of the Dirac equation (\ref{Eq: The Dirac eq in curved spacetime}) could be done through several equivalent approaches \cite{PhysRevLett.31.1265,Chandrasekhar:1976ap,PhysRevD.19.1093,Chandrasekhar:1984siy}. Here, following \cite{0264-9381-32-18-184001,Carter:1968ks}, we employ the canonical orthonormal (symmetric) tetrad for the RN spacetime, and decompose the Dirac four-spinor as
\begin{equation}\label{Eq: ansatz}
	\bm{\Psi}=\frac{1}{\sqrt{r\Delta^{1/2}}}\left(\begin{array}{c}
	\psi_-(r,\theta)\\
	\psi_+(r,\theta)
	\end{array}\right)e^{-i\omega t+im\phi},
\end{equation}
where
\begin{equation}
	\psi_+=\left(\begin{array}{c}
	R_1(r)S_1(\theta)\\
	R_2(r)S_2(\theta)
	\end{array}\right),\;\;\mathrm{and}\;\;\psi_-=-\left(\begin{array}{c}
	R_2(r)S_1(\theta)\\
	R_1(r)S_2(\theta)
	\end{array}\right).
\end{equation}
Substituting the ansatz (\ref{Eq: ansatz}) and the background metric (\ref{Eq: the line element}) into the Dirac equation (\ref{Eq: The Dirac eq in curved spacetime}) yields the following coupled first-order differential equations
\begin{equation}\label{Eq: R_1}
	\sqrt{\Delta}\left(\frac{d}{dr}-\frac{iK}{\Delta}\right)R_1=\left(\lambda+i\mu r\right)R_2,
\end{equation}
\begin{equation}\label{Eq: R_2}
	\sqrt{\Delta}\left(\frac{d}{dr}+\frac{iK}{\Delta}\right)R_2=\left(\lambda-i\mu r\right)R_1,
\end{equation}
and
\begin{equation}\label{Eq: S_1}
	\left(\frac{d}{d\theta}+\frac{1}{2}\cot\theta-m\csc\theta\right)S_1=+\lambda S_2,
\end{equation}
\begin{equation}\label{Eq: S_2}
	\left(\frac{d}{d\theta}+\frac{1}{2}\cot\theta+m\csc\theta\right)S_2=-\lambda S_1,
\end{equation}
where $K=\omega r^2-qQr$ and $\lambda$ is the separation constant. 

Solutions of the angular Eqs.(\ref{Eq: S_1}) and (\ref{Eq: S_2}) are spin-weighted spherical harmonics and the eigenvalues are given by (See \cite{0264-9381-26-17-175020} for details)
\begin{equation}\label{Eq: lambda and ell}
	\lambda=\left\{\
	\begin{aligned}
	&+\ell,\;\;\;\;\;\;\;\;\mathrm{for}\;\;\ell=j+\frac{1}{2},\\&-1-\ell,\;\;\mathrm{for}\;\;\ell=j-\frac{1}{2},
\end{aligned}
\right.
\end{equation}
where $j$ is a positive half-integer, i.e. $j=1/2,3/2,\cdots$, which denotes the total angular momentum. The orbital angular momentum is characterized by the non-negative integer $\ell=0,1,2,\cdots$. By the way, the relation between $\lambda$ and $j$ is given by
\begin{equation}\label{Eq: lambda and j}
	|\lambda|=j+\frac{1}{2}.
\end{equation}
It should be noted that Eqs.(\ref{Eq: R_1}) and (\ref{Eq: R_2}) can be written as the following second-order form\footnote{The corresponding second-order equation of $R_2$ is similar to this equation.}
\begin{equation}\label{Eq: 2nd order radial equation}
\sqrt{\Delta}\frac{d}{dr}\left(\sqrt{\Delta}\frac{dR_1}{dr}\right)-\frac{i\mu\Delta}{\lambda+i\mu r}\frac{dR_1}{dr}+UR_1=0,
\end{equation}
where
\begin{widetext}
	\begin{equation}
	U=\frac{K^2+i(r-M)K}{\Delta}-2i\omega r+iqQ-\frac{\mu K}{\lambda+i\mu r}-\lambda^2-\mu^2r^2.
	\end{equation}
\end{widetext}
This second-order equation can further be rewritten in the form of a Schr\"{o}dinger-like wave equation
\begin{equation}\label{Eq: 2nd order radial equation for psi}
\frac{d^2\psi}{dr^2}+\left(\omega^2-V\right)\psi=0,
\end{equation}
where the new radial function $\psi$ is defined by
\begin{equation}
\psi(r)=F^{-1}R_1(r),
\end{equation}
and the effective potential is given by
\begin{equation}\label{Eq: effective potential}
V=\omega^2-\frac{U}{\Delta}-\frac{F''}{F}-\left(\frac{\Delta'}{2\Delta}-\frac{i\mu}{\lambda+i\mu r}\right)\frac{F'}{F}.
\end{equation}
Here, the auxiliary function $F(r)$ is defined by
\begin{equation}
F(r)=\left(1+\frac{i\mu r}{\lambda}\right)^{\frac{1}{2}}\Delta^{-\frac{1}{4}}.
\end{equation}
\subsection{Quasibound states}
As is known, quasibound state solutions are ingoing at the horizon, and exponentially decaying at infinity. To obtain such solutions, we solve the radial equations (\ref{Eq: R_1}) and (\ref{Eq: R_2}) under appropriate boundary conditions. By doing this, we can pick out a discrete set of complex frequencies, which correspond to the quasibound states of the Dirac field. Here, we express a complex frequency by $\omega=\omega_R+i\omega_I$. In the limit of small coupling parameters,
\begin{equation}\label{cd: small parameter}
\mathcal{O}(|qQ|)=\mathcal{O}(M\mu)\equiv\mathcal{O}(\epsilon),\;\;\epsilon\ll1,
\end{equation}
the massive charged Dirac field in the RN BH spacetime has the following spectrum \cite{Ternov1980}
\begin{equation}\label{Eq: analytic results}
	1-\frac{\omega_R}{\mu}\simeq\frac{\left(M\mu-qQ\right)^2}{2\tilde{n}^2}+\mathcal{O}\left(\epsilon^4\right),
\end{equation}
where the principal quantum number
\begin{equation}\label{Eq: principal quantum number}
	\tilde{n}=\left\{\
	\begin{aligned}
	&n+\ell,\;\;\;\;\;\;\;\;\mathrm{for}\;\;\ell=j+\frac{1}{2},\\&n+\ell+1,\;\;\mathrm{for}\;\;\ell=j-\frac{1}{2},
	\end{aligned}
	\right.
\end{equation}
and the excitation number $n=0,\;1,\;2,\;\cdots$. From Eq.(\ref{Eq: analytic results}), the binding energy $\omega_R-\mu$ tends to zero in the  limit of $M\mu\rightarrow qQ$. One may also find that Eq.(\ref{Eq: principal quantum number}) can be written in a simpler form
\begin{equation}
	\tilde{n}=n+j+\frac{1}{2}.
\end{equation}

In next section, we shall use the continued fraction method \cite{Leaver285,PhysRevD.41.2986} to calculate the quasibound state frequencies beyond the small parameter regime (\ref{cd: small parameter}), where the analytic formula (\ref{Eq: analytic results}) is no longer valid. It is worth noting that, due to the absence of superradiance, no growing modes could be found for Dirac fields around a RN BH.

\section{Frequency domain analysis}\label{Sec: Frequency domain method}
The main problem for solving the coupled equations (\ref{Eq: R_1}) and (\ref{Eq: R_2}) via the continued fraction method is that the asymptotic solutions of $R_1$ and $R_2$ are different at $r_+$ and infinity. Accordingly, we define two new radial functions as
\begin{equation}
	R_2=\frac{\sqrt{\Delta}}{r-r_+}R_+,\;\;\mathrm{and}\;\;R_1=R_-,
\end{equation}
then, the radial equations (\ref{Eq: R_1}) and (\ref{Eq: R_2}) become
\begin{equation}\label{Eq: R_-}
	\left(\frac{d}{dr}-\frac{iK}{\Delta}\right)R_-=\frac{\lambda+i\mu r}{r-r_+}R_+,
\end{equation}
and
\begin{equation}\label{Eq: R_+}
	\left(\frac{d}{dr}+\frac{iK+r-M}{\Delta}-\frac{1}{r-r_+}\right)R_+=\frac{\lambda-i\mu r}{r-r_-}R_-.
\end{equation}
Close to $r_+$, the "ingoing" solutions behave as
\begin{equation}
	\lim\limits_{r\rightarrow r_+}R_\pm\sim\left(r-r_+\right)^\rho
\end{equation}
where
\begin{equation}
	\rho=\frac{1}{2}-\frac{ir^2_+(\omega-\omega_c)}{r_+-r_-},\;\;\mathrm{and}\;\;\omega_c=\frac{qQ}{r_+}.
\end{equation}
At infinity, the asymptotic solutions of Eqs.(\ref{Eq: R_-}) and (\ref{Eq: R_+}) are
\begin{equation}\label{Eq: behavior at infinity}
	\lim\limits_{r\rightarrow \infty}R_\pm\sim r^\chi e^{kr},
\end{equation}
where
\begin{equation}
	k=\pm\sqrt{\mu^2-\omega^2},
\end{equation}
and
\begin{equation}
	\chi=\frac{M\left(\mu^2-2\omega^2\right)+qQ\omega}{k}.
\end{equation}
Here, we are only interested in quasibound states which decay exponentially at infinity. Thus, we choose $\mathrm{Re}(k)<0$.

Based on the discussions above, solutions of Eqs.(\ref{Eq: R_-}) and (\ref{Eq: R_+}) can be expanded as
\begin{equation}\label{Eq: ansatz for series_levear}
	\left(\begin{array}{c}
	R_-\\
	R_+
	\end{array}\right)=\left(r-r_-\right)^\chi e^{kr}\sum_{n=0}^{\infty}\bm{U}_n\left(\frac{r-r_+}{r-r_-}\right)^{n+\rho},
\end{equation}
where $\bm{U}_n$ are two dimensional vectorial coefficients. Substituting the series (\ref{Eq: ansatz for series_levear}) into Eqs.(\ref{Eq: R_-}) and (\ref{Eq: R_+}) leads to a three-term matrix-valued recurrence relation,
\begin{equation}
	\bm{\alpha}_0\bm{U}_1+\bm{\beta}_0\bm{U}_0=0,
\end{equation}
\begin{equation}
	\bm{\alpha}_n\bm{U}_{n+1}+\bm{\beta}_n\bm{U}_n+\bm{\gamma}_n\bm{U}_{n-1}=0,\;\;\;\;n>0.
\end{equation}
Here, $\bm{\alpha}_n$, $\bm{\beta}_n$, $\bm{\gamma}_n$ are all $2\times2$ matrices and can be expressed, respectively, as
\begin{equation}
	\bm{\alpha}_n=\left(\begin{array}{c c}
	n+\frac{1}{2}+2\rho\;\;\;\;& -\lambda-i\mu r_+\\\\
	0 & n+1
	\end{array}\right),
\end{equation}

\begin{equation}
	\bm{\beta}_n=\left(\begin{array}{c c}
	-2n+c_0\;\;\;\;& \lambda+i\mu r_-\\\\
	-\lambda+i\mu r_+ & -2n+c_0+c_1
	\end{array}\right),
\end{equation}

\begin{equation}
	\bm{\gamma}_n=\left(\begin{array}{c c}
	n-c_0+c_2\;\;\;\;& 0\\\\
	\lambda-i\mu r_- & n-c_0+c_2+c_3
	\end{array}\right),
\end{equation}
where $c_0$, $c_1$, $c_2$ and $c_3$ are given by
\begin{equation}
	\begin{aligned}
	c_0=&\frac{1}{2}-3\rho-i\omega r_++\frac{k(3r_+-r_-)}{2}-\\&\frac{r_-}{r_+}\left(\rho-\frac{1}{2}\right)+\frac{(i-2i\rho+\omega r_+)\omega}{2k}\left(1-\frac{r_-}{r_+}\right),
	\end{aligned}
\end{equation}
\begin{equation}
	c_1=2\left(\rho+i\omega r_+\right)+\frac{r_-}{r_+}\left(2\rho-1\right),
\end{equation}
\begin{equation}
	c_2=-\frac{1}{2}-2\rho+(r_+-r_-)(k-i\omega),
\end{equation}
\begin{equation}
	c_3=-\frac{1}{2}+\frac{r_-(1-2\rho-2i\omega r_+)}{r_+}.
\end{equation}
The quasibound frequencies are roots of equation $\bm{M}_0\bm{U}_0=0$, where
\begin{equation}
	\bm{M}_0=\bm{\beta}_0+\bm{\alpha}_0\bm{R}_0,
\end{equation}
with $\bm{U}_{n+1}=\bm{R}_n\bm{U}_n$ and
\begin{equation}\label{Eq: relation between R_n's}
	\bm{R}_n=-\left(\bm{\beta}_{n+1}+\bm{\alpha}_{n+1}\bm{R}_{n+1}\right)^{-1}\bm{\gamma}_{n+1}.
\end{equation}
Then, the nontrivial solutions $\bm{U}_0$ exist if
\begin{equation}\label{Eq: det M_0}
	\det|\bm{M}_0|=0.
\end{equation}

\begin{figure}
	\includegraphics[width=0.5\textwidth]{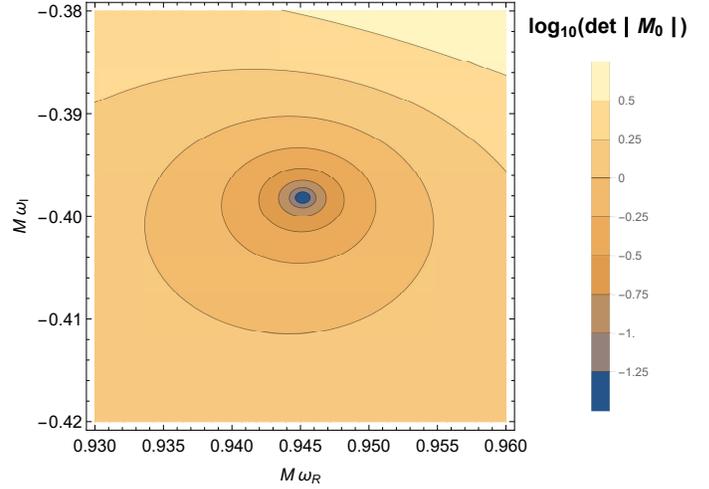}
	\caption{Contour plot of the logarithm of the magnitude of the determinant of matrix $\bm{M}_0$ as a function of the real and imaginary parts of the frequency. Here, we set $Q=0.5M,\;qM=-0.4,\;M\mu=1$ and $\lambda=-1$.}
	\label{fig: Contour plot} 
\end{figure}

For physical accepted solutions, the series expansion (\ref{Eq: ansatz for series_levear}) should converge at spatial infinity. Thus, we may fix a large truncation order $N$ and initialize $\bm{R}_N$ as an identity matrix\footnote{ Equivalently, we require $\bm{U}_{N+1}=\bm{U}_N$.}. Then, Eq.(\ref{Eq: relation between R_n's}) can be used to construct the matrix $\bm{M}_0$ (See \cite{doi:10.1142/S0217751X13400186} for more details). Finally, the quasibound state frequencies are the roots of Eq.(\ref{Eq: det M_0}), which correspond to the local minimum points of $\det|\bm{M}_0|$ in a plane spanned by $\omega_R$ and $\omega_I$. Fig.\ref{fig: Contour plot} shows $\det|\bm{M}_0|$ as a function of the real and imaginary parts of the frequency for $\ell=0,\;j=\frac{1}{2}$. To obtain the quasibound state frequencies we use a 2D root-finding algorithm. Using this approach, one may easily find that the local minimum in the Fig.\ref{fig: Contour plot} gives $M\omega=0.945159-0.398158i$, which is the frequency of the fundamental mode.

\begin{figure*}    
	\subfigure{\includegraphics[width=0.4\textwidth]{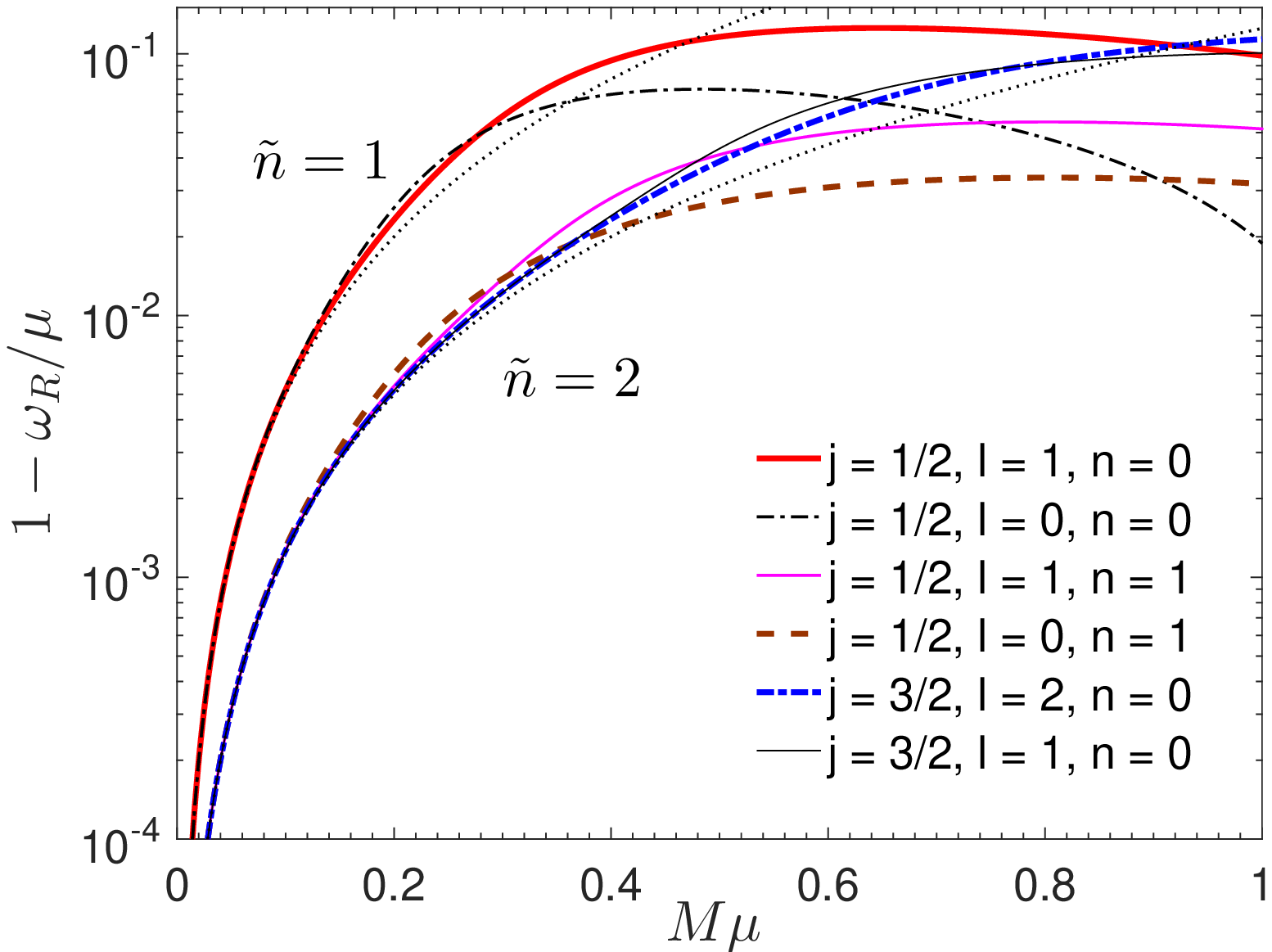}}
	\subfigure{\includegraphics[width=0.4\textwidth]{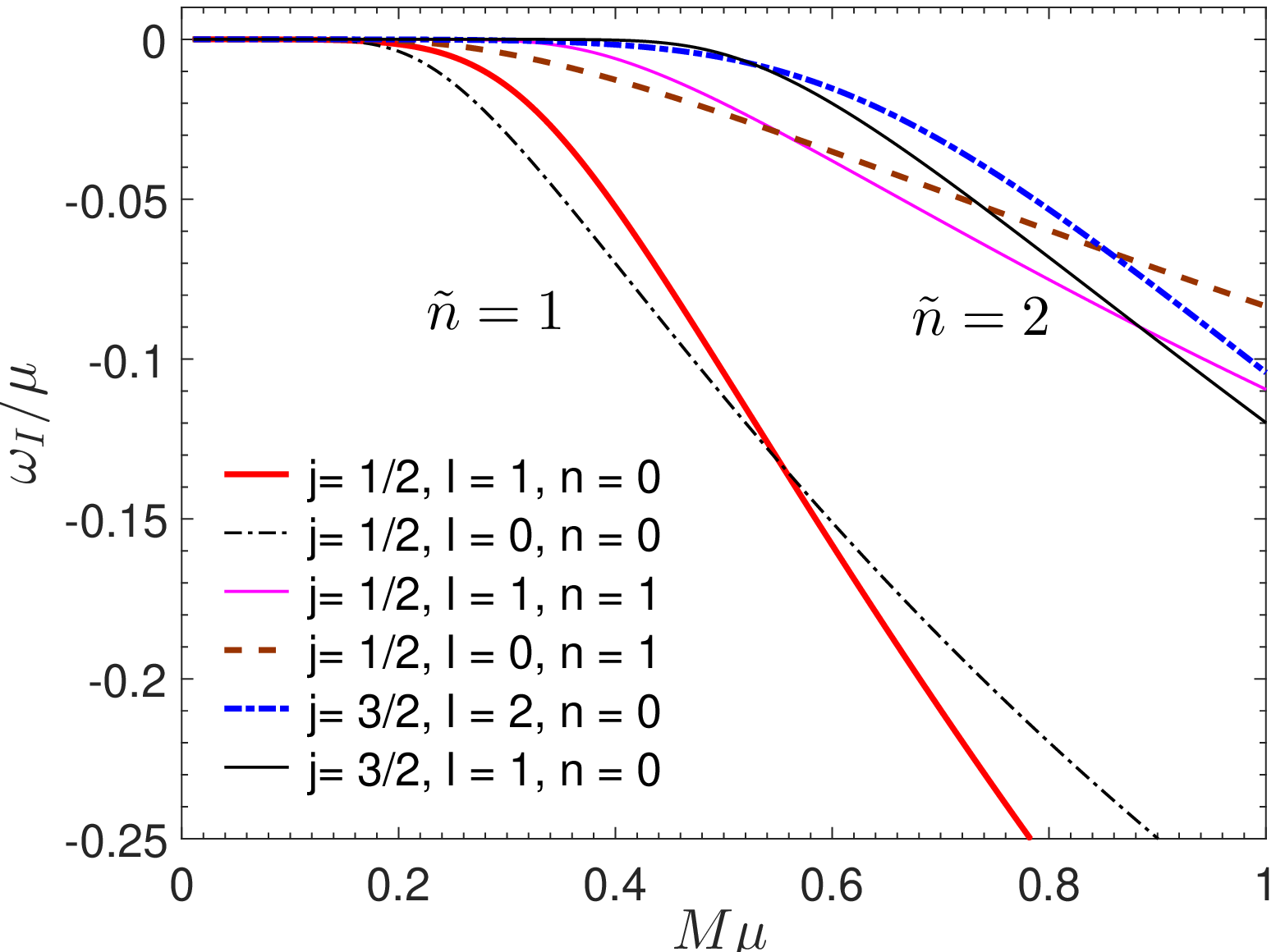}}
	\caption{The quasibound state frequencies, with $q=Q=0$, as a function of field mass $\mu$. In the left panel, the top and bottom branches correspond to $\tilde{n}=1$ ($j=1/2,\;n=0$) and $\tilde{n}=2$ ($j=1/2,\;n=1$ or $j=3/2,\;n=0$) modes, respectively. Here, two dotted black lines (Obtained from the analytic formula Eq.(\ref{Eq: analytic results})) are also presented. Quasibound states with the same values of $\tilde{n}$ degenerate in the small mass limit $M\mu\ll1$. The right panel shows the corresponding value of $\omega_I$ as a function of $M\mu$.}
	\label{fig: neutral}
\end{figure*}

Due to the spin-orbit coupling, Dirac fields around a BH have a rich spectrum, even for the case in which both the field and BH are neutral. Fig.\ref{fig: neutral} shows the real part $\omega_R$ (expressed by $1-\omega_R/\mu$ in the left panel) and imaginary part $\omega_I$ (the right panel) of the first few modes, as functions of $M\mu$, for Dirac fields in the Schwarzschild spacetime ($Q=0$). In Table \ref{table:2} we present some reference frequencies for initial estimations of the root-finding algorithm.

\begin{table}
	\caption{A list of the frequencies of  selected  quasibound states of the Dirac field with $q=Q=0$ and $M\mu=0.39$.}\label{table:2}
	\begin{tabular}{ccccc}\hline\hline
		\centering
		$\tilde{n}$\;\;\;\;\;& $n$ \;\;\;\;\;& $j$ \;\;\;\;\;& $\ell$\;\;\;\;\;&$M\omega$\\
		$1$        \;\;\;\;\;& $0$ \;\;\;\;\;& $1/2$ \;\;\;\;\;& $0$ \;\;\;\;\;&$0.36308-0.02568i$\\
		\;\;\;\;\;& $0$ \;\;\;\;\;& $1/2$ \;\;\;\;\;& $1$ \;\;\;\;\;&$0.35447-0.01862i$\\
		$2$        \;\;\;\;\;& $1$ \;\;\;\;\;& $1/2$ \;\;\;\;\;& $0$ \;\;\;\;\;&$0.38195-0.00457i$\\
		\;\;\;\;\;& $1$ \;\;\;\;\;& $1/2$ \;\;\;\;\;& $1$ \;\;\;\;\;&$0.37964-0.00196i$\\
		\;\;\;\;\;& $0$ \;\;\;\;\;& $3/2$ \;\;\;\;\;& $1$ \;\;\;\;\;&$0.38141-0.00057i$\\
		\;\;\;\;\;& $0$ \;\;\;\;\;& $3/2$ \;\;\;\;\;& $2$ \;\;\;\;\;&$0.38121-0.00006i$\\
		\hline\hline
	\end{tabular}
\end{table}

From the left panel of Fig.\ref{fig: neutral}, we see that for small values of $M\mu$, the spectrum of modes with the same value of $\tilde{n}=n+j+\frac{1}{2}$ are indistinguishable, and they are in good agreement with the analytic formula Eq.(\ref{Eq: analytic results}) (denoted by black dotted lines in the plot). As $M\mu$ increases, the spin-orbit coupling begins to affect the quasibound spectrum, resulting in a removal of the degeneracy of the same value of $\tilde{n}$. The dependence of $\omega_I(\mu)$ is simpler. The right panel of Fig.\ref{fig: neutral} shows that the absolute value of $\omega_I$ decreases as $M\mu$ decreases. In the limit of $M\mu\rightarrow0$, the value of $|\omega_I|$ also tends to zero, which means that ultralight Dirac field could be arbitrarily long-lived around a Schwarzschild BH.

We now turn to the case in which both the BH and the Dirac field are electrically charged. Without loss of generality, we assume that the charge of BH is always positive $Q>0$. We shall show how the electromagnetic interaction changes the spectrum and how long-lived modes are achieved.

\begin{figure*}
	\subfigure{
		\includegraphics[width=0.4\textwidth]{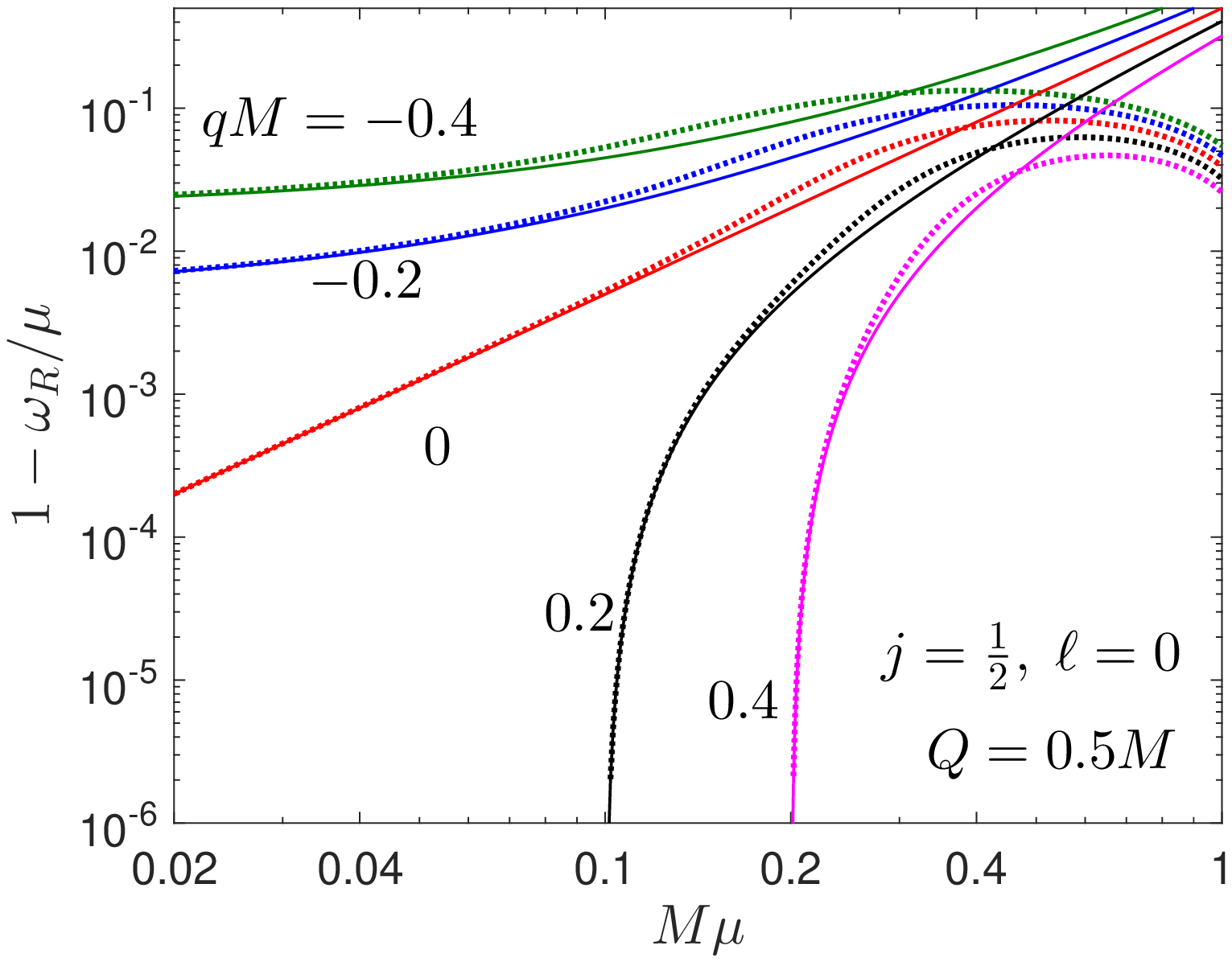}
	}
	\subfigure{
		\includegraphics[width=0.4\textwidth]{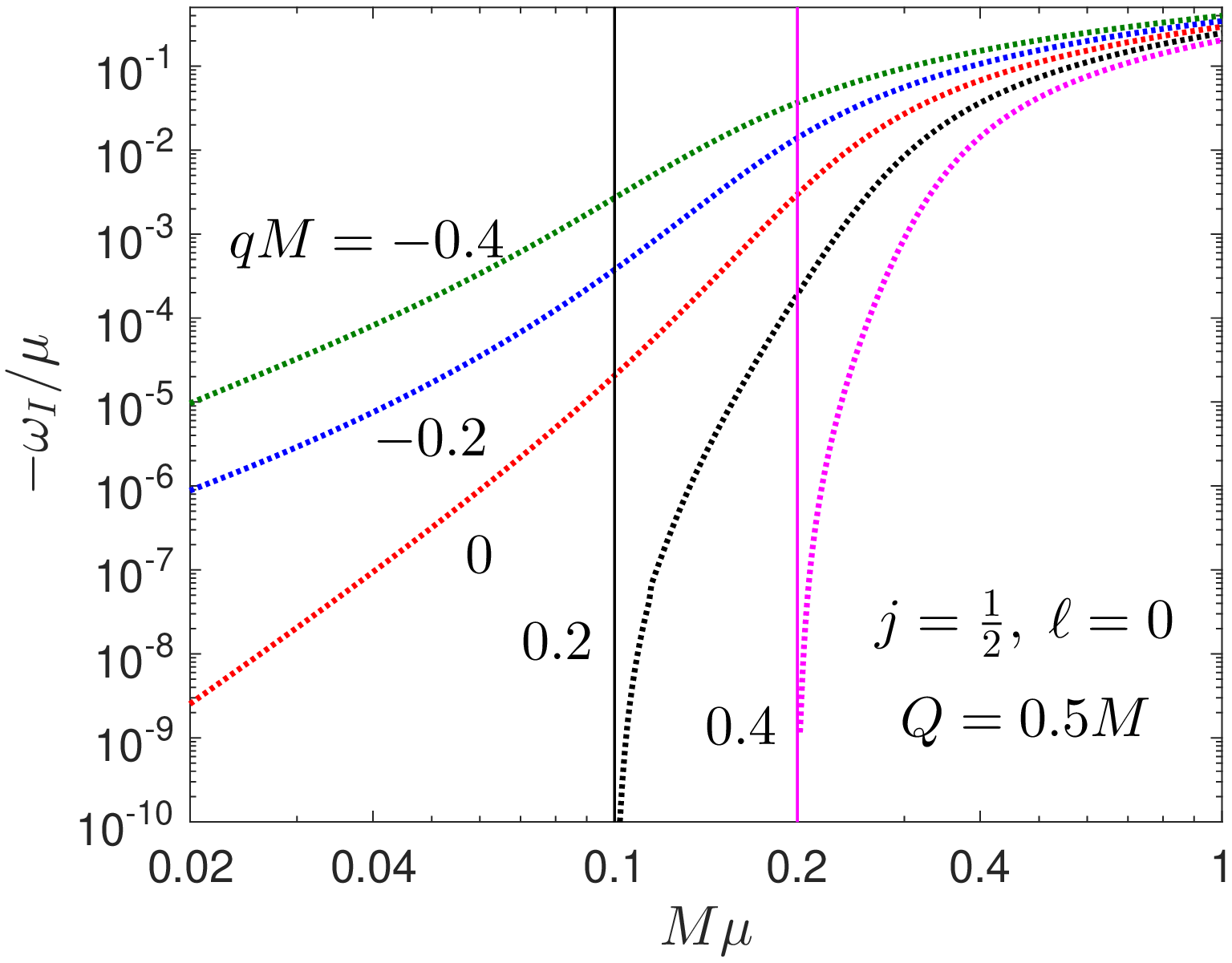}
	}
	\caption{The lowest quasibound state frequencies, with $j=1/2$ and $n=\ell=0$, as a function of $M\mu$ for various values of charge $q$ of the Dirac field. The left panel shows the real part $\omega_R$, whereas the right panel shows the corresponding lines for the imaginary part $\omega_I$. The solid lines in the left panel correspond to Eq.(\ref{Eq: analytic results}). The vertical lines in the right panel give, for the corresponding value of $q$, $M\mu=qQ$.}
	\label{fig: long-lived Dirac cloud}
\end{figure*}

Fig.\ref{fig: long-lived Dirac cloud} shows the effect of field charge $q$ on the quasibound state spectrum. Clearly, more negative field charge corresponds to smaller values of $\omega_R$ but larger values of $|\omega_I|$, which suggests that for positive BH charge, Dirac field with negative charge decays faster than the one with positive charge. More interestingly, for positively charged Dirac field, the value of $M\mu$ of the quasibound states is bounded below by $M\mu_\mathrm{min}=qQ$, which is denoted by the vertical lines in the right panel of Fig.\ref{fig: long-lived Dirac cloud}. By decreasing $M\mu$ close to its minimum value, i.e., $M\mu\gtrsim qQ$, one gets very long-lived modes since the decay rate $|\omega_I|$ tends to zero rapidly.

\section{Time domain analysis}\label{Sec: Time domain method}
Massive test fields around a BH can form both quasinormal modes (QNMs) and quasibound states. Generally speaking, QNMs would appear in early times and damp quickly, whereas quasibound states decay on larger time scales. It follows that given generic initial data, quasibound states would dominate the waveform of massive fields around BHs. This is well understood for scalar and Proca fields \cite{PhysRevD.84.083008,PhysRevD.90.065019,PhysRevD.87.124026,PhysRevD.87.043513,PhysRevD.89.083006,Yoshino:2012kn,Yoshino:2013ofa,Yoshino:2014wwa,PhysRevD.89.063005,Degollado:2009rw,PhysRevD.93.104011}. Unfortunately, studies of the time evolution for massive Dirac fields are relatively few \cite{0264-9381-32-18-184001,PhysRevD.89.043006,PhysRevD.72.027501}.

Here, we consider the time evolution of massive charged Dirac fields around a RN BH. From Eqs.(\ref{Eq: R_1}) and (\ref{Eq: R_2}), the Dirac wave equations in the time domain can be written in the form
\begin{equation}
	\partial_tR_1+\partial_xR_1+i\frac{qQ}{r}R_1=\frac{\sqrt{\Delta}}{r}\left(i\mu+\frac{\lambda}{r}\right)R_2,
\end{equation}
\begin{equation}
	\partial_tR_2-\partial_xR_2+i\frac{qQ}{r}R_2=\frac{\sqrt{\Delta}}{r}\left(i\mu-\frac{\lambda}{r}\right)R_1,
\end{equation}
where the tortoise coordinate $x=\int dr\;r^2/\Delta$. We use the method of lines with a second-order finite difference stencil for the spatial derivatives, and third-order Runge–Kutta integrator for the time integration. To minimize the boundary reflections, we place the inner boundary very close to the event horizon, typically $x=-1000M$, and put the outer boundary far away from the BH; furthermore, we suppose that solutions are continuous enough and extrapolate values of $R_1$ and $R_2$ at boundary points from the inner points at each time step. In addition, we include a Kreiss-Oliger dissipation to ensure the numerical stability. We choose Gaussian wave packet
\begin{equation}\label{Eq: initial wave packet}
R_1(0,x)=e^{-\left(x-x_g\right)^2/2\sigma^2},\;\;R_2(0,x)=0,
\end{equation}
as the initial data. The time derivatives of $R_1$ and $R_2$ are set to zero initially. For convenience, we set $r_+=1$ and measure all quantities in terms of $r_+$ in the time evolution of the Dirac field.

\begin{figure*}    
	\subfigure{\includegraphics[width=0.35\textwidth]{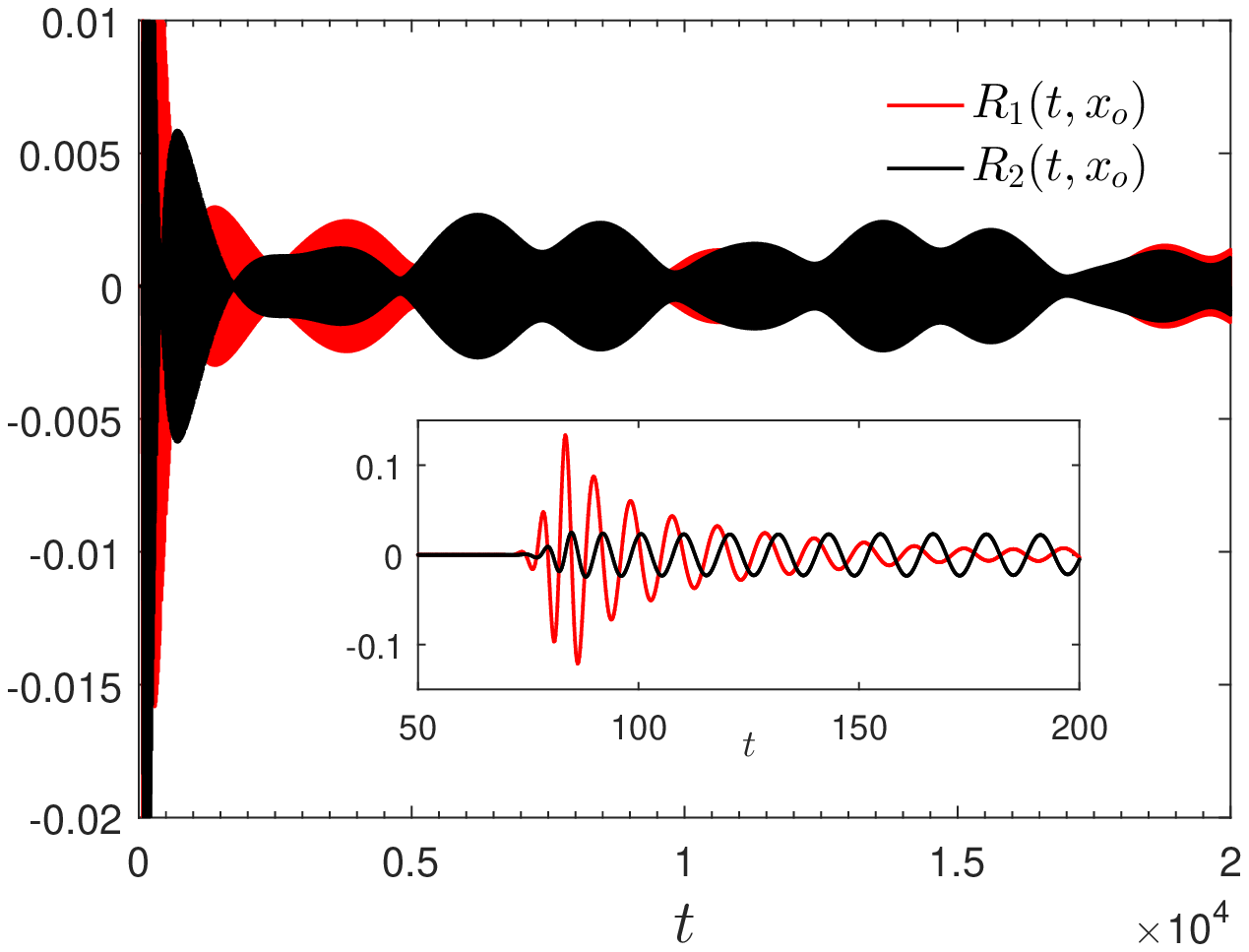}}
	\subfigure{\includegraphics[width=0.35\textwidth]{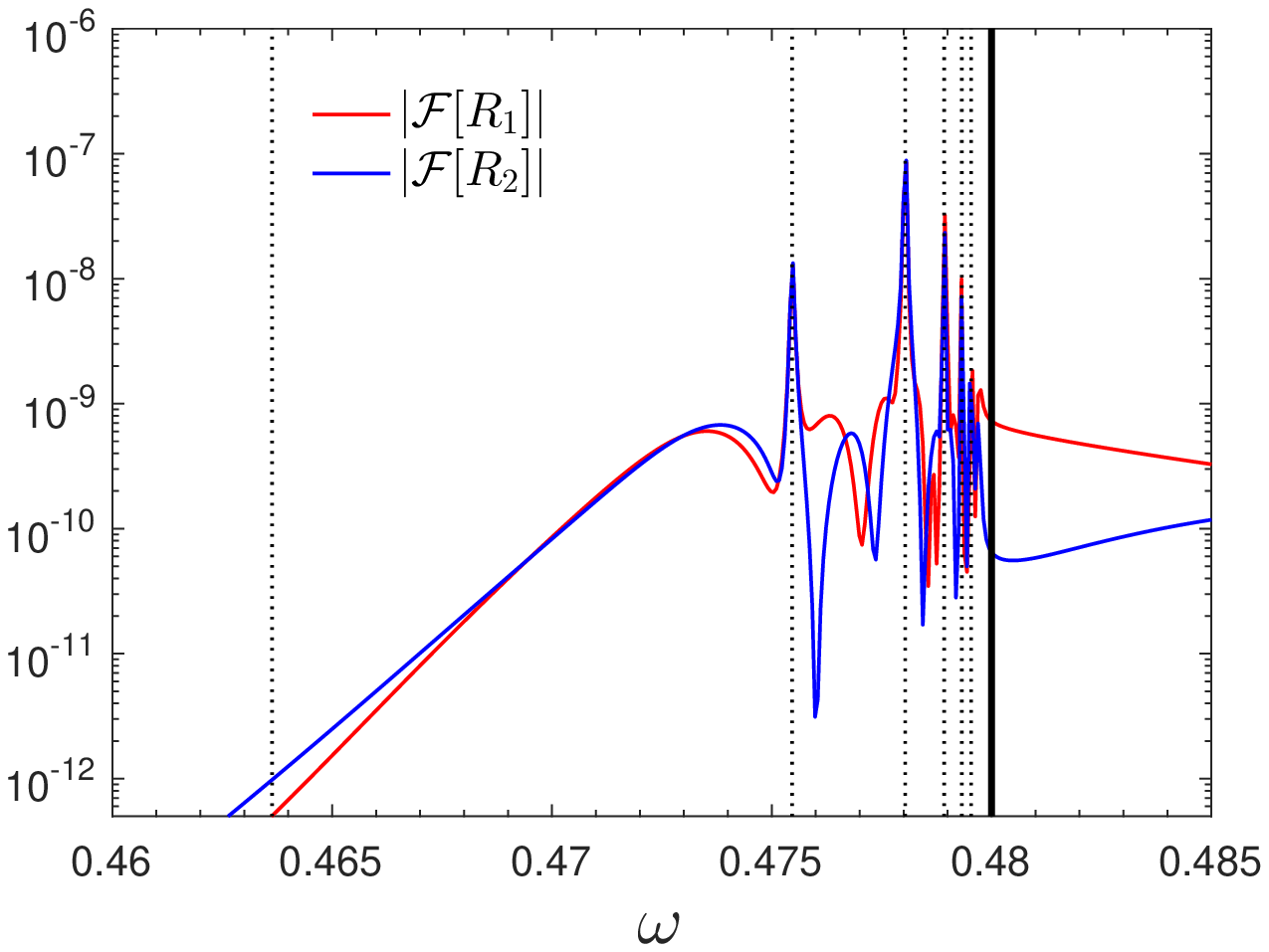}}
	\subfigure{\includegraphics[width=0.35\textwidth]{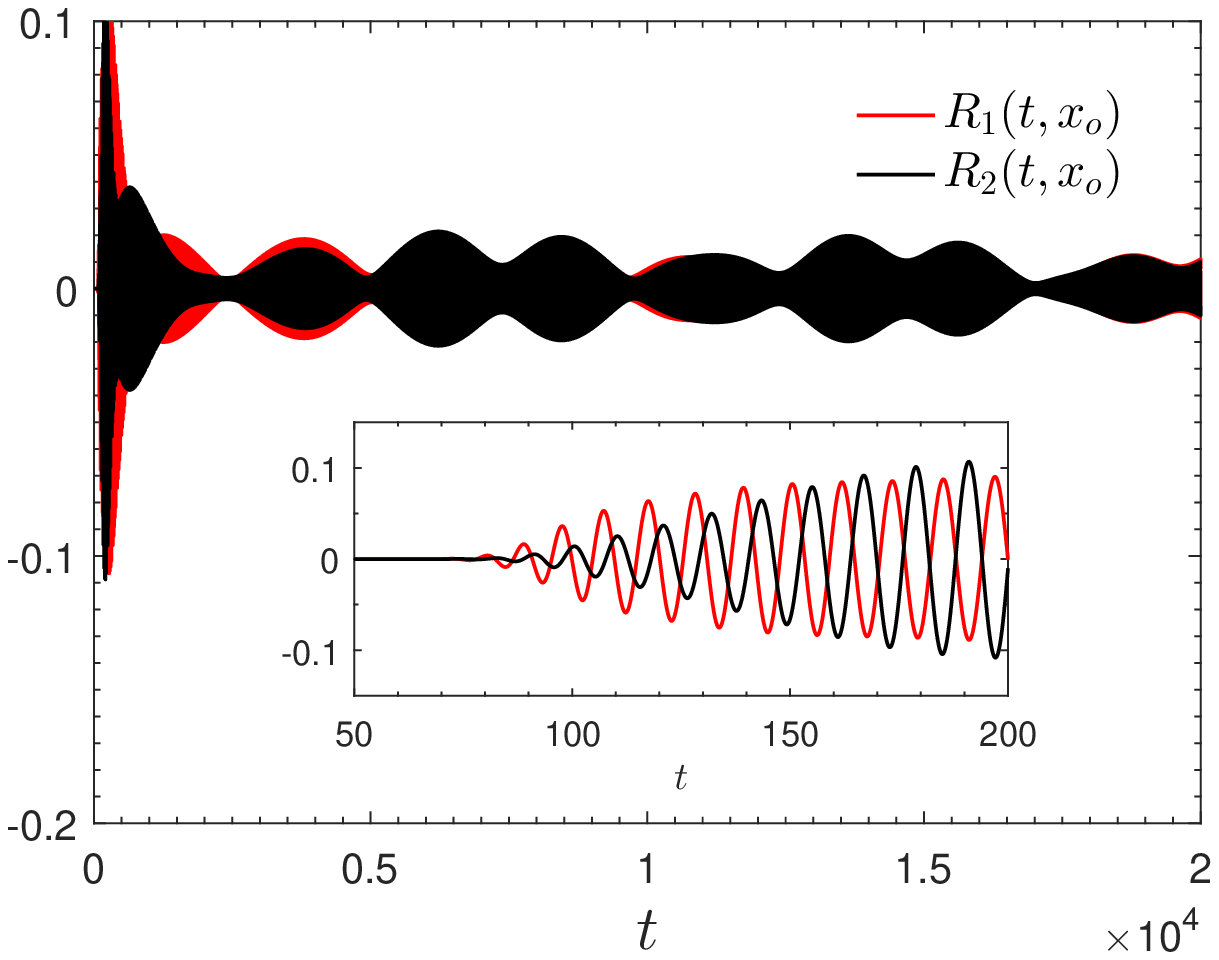}}
	\subfigure{\includegraphics[width=0.35\textwidth]{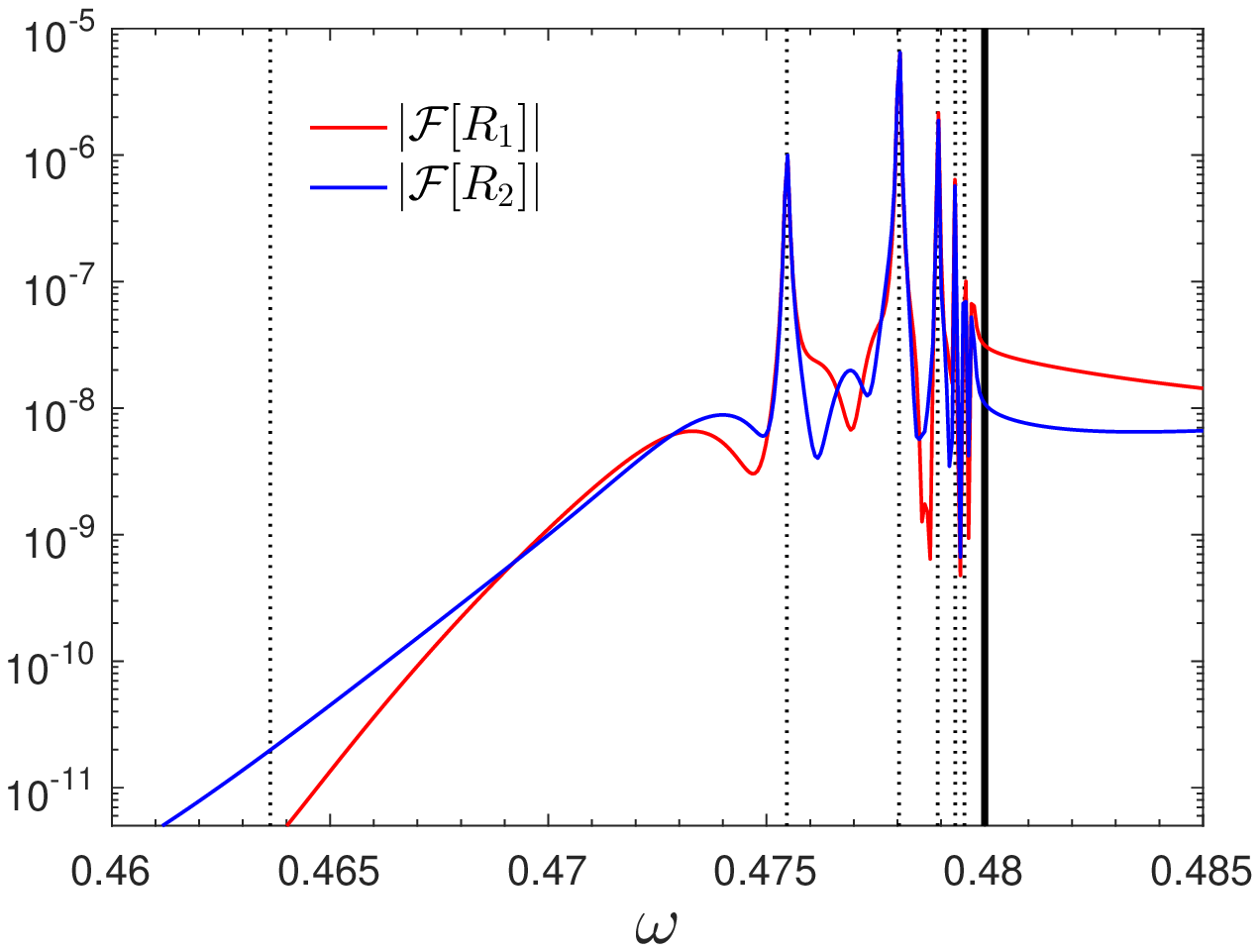}}
	\subfigure{\includegraphics[width=0.35\textwidth]{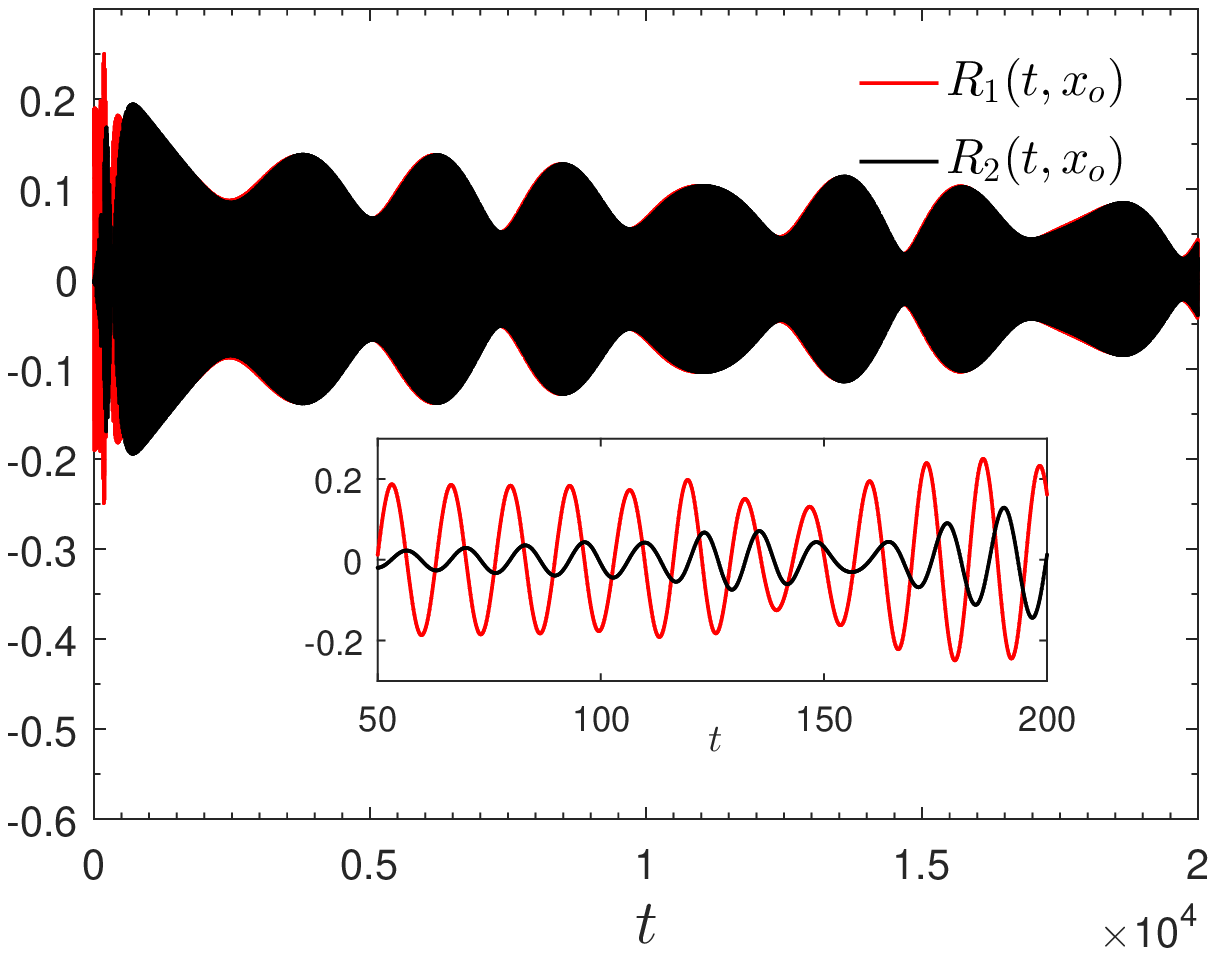}}
	\subfigure{\includegraphics[width=0.35\textwidth]{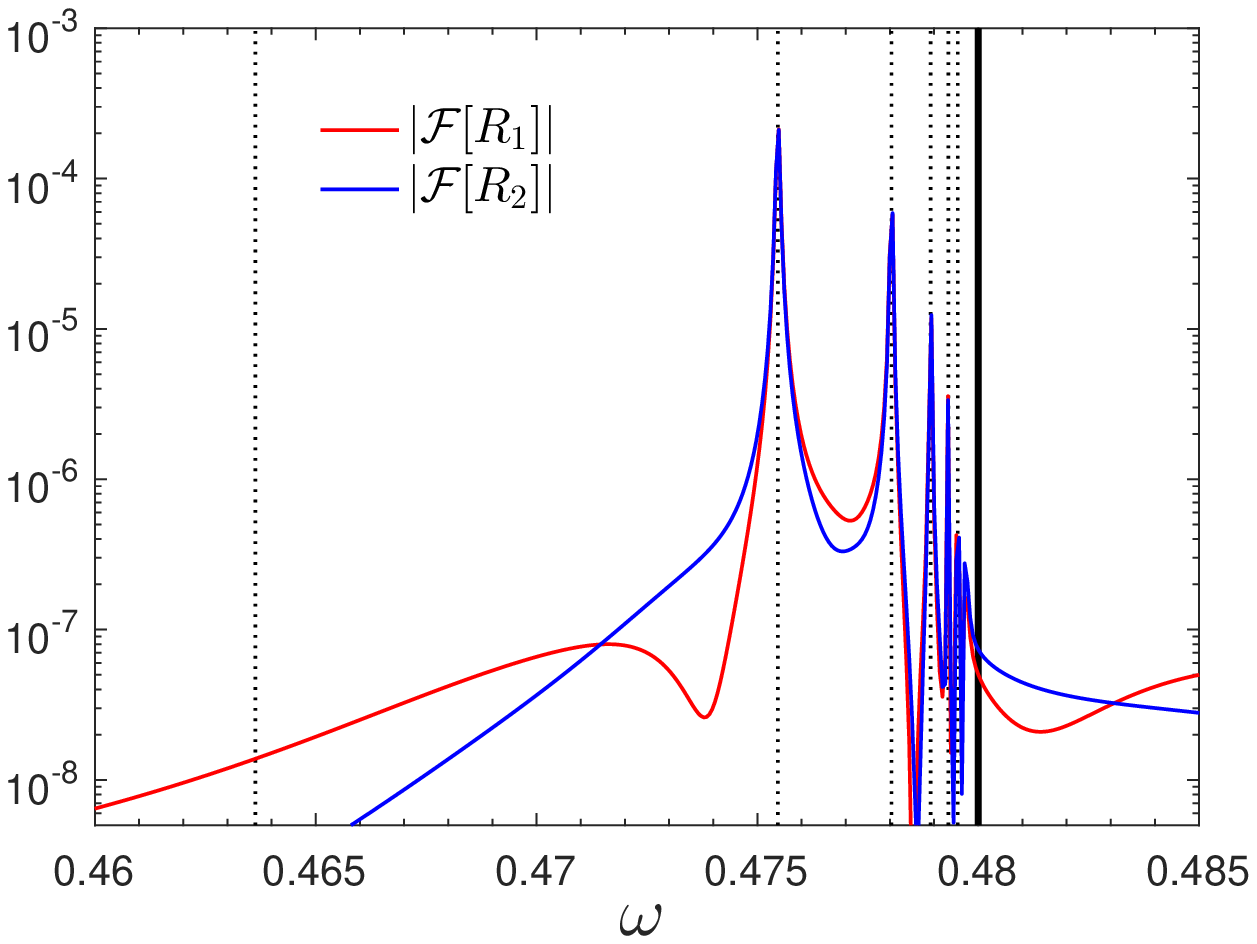}}
	\caption{Left panels: Time evolution of the real parts of $R_1$ and $R_2$ observed at $x_o=80$ for initial data of the form in Eq.(\ref{Eq: initial wave packet}), with $\lambda=1$, $r_-=0.8$, $q=0.2$ and $\mu=0.48$. Parameters of the initial wave packet in Eq.(\ref{Eq: initial wave packet}) are given by $x_g=7$ and $\sigma=\left\lbrace0.4,\;4,\;40\right\rbrace $, from top to bottom. Right panels: Corresponding Fourier spectra of the evolutions in the left panels, up to $t=10^5$. The Dashed vertical lines in the right panels indicate the frequencies of the quasibound states computed via the frequency-domain method discussed in previous section, and Table \ref{table:1} is a collection of these results. The thick vertical lines give $\omega=\mu$, which denotes the upper bound of the frequency of the quasibound states.}
	\label{fig: time-domain}
\end{figure*}

The typical results for the time evolution are presented in Fig.\ref{fig: time-domain}, where we plot the waveforms of $R_1$ and $R_2$ at a fixed point $x_o=80$ for different values of $\sigma$ in the left panels, and plot the corresponding Fourier spectra in the right panels. The left panels show the beating effect of the Dirac field, which has been found for bosonic fields in Kerr spacetime \cite{PhysRevD.87.043513}. Although the evolutions of $R_1$ and $R_2$ are different at early times and depend strongly on the initial data we choose (See the insets in the left panels of Fig.\ref{fig: time-domain}), the beating patterns of them are quite similar at late times of the evolution.

As pointed out in \cite{PhysRevD.87.043513}, such beating effect results from the interference between several modes with different frequencies. Indeed, from the right panels of Fig.\ref{fig: time-domain}, we see that there are several visible peaks in the Fourier spectra which correspond to the dominating modes at late time stages of the evolutions. These peaks coincide with the quasibound state frequencies. Thus we conclude that quasibound states of Dirac fields could be excited by a generic Gaussian initial data and would dominate the waveforms of long-timescale evolutions. However, one may easily observe that the $n=0$ state is missing in the Fourier spectrum. This is  reasonable because the absolute value of $\omega_I$ of the $n=0$ state is about two orders of magnitude larger than those of higher overtones, which makes the $n=0$ mode decay rapidly and disappear in the Fourier spectrum.

\begin{table}
	\caption{The first few quasibound state frequencies computed via the continued fraction method. The Dirac field and BH parameters are the same as those in Fig.\ref{fig: time-domain}.}\label{table:1}
	\begin{tabular}{cc}\hline\hline
		\centering
		$n$ \;\;\;\;\;\;\;\; & $ \omega\;r_+$\\
		$0$ \;\;\;\;\;\;\;\; & $ 0.463633 - 1.044\times10^{-3} i $\\
		$1$ \;\;\;\;\;\;\;\; & $ 0.475466 - 3.899\times10^{-5} i $\\
		$2$ \;\;\;\;\;\;\;\; & $ 0.478036 - 1.507\times10^{-5} i $\\
		$3$ \;\;\;\;\;\;\;\; & $ 0.478923 - 6.685\times10^{-6} i $\\
		$4$ \;\;\;\;\;\;\;\; & $ 0.479323 - 3.429\times10^{-6} i $\\
		$5$ \;\;\;\;\;\;\;\; & $ 0.479536 - 1.955\times10^{-6} i $\\
		\hline\hline
	\end{tabular}
\end{table}

\begin{figure*}    
	\subfigure{\includegraphics[width=0.4\textwidth]{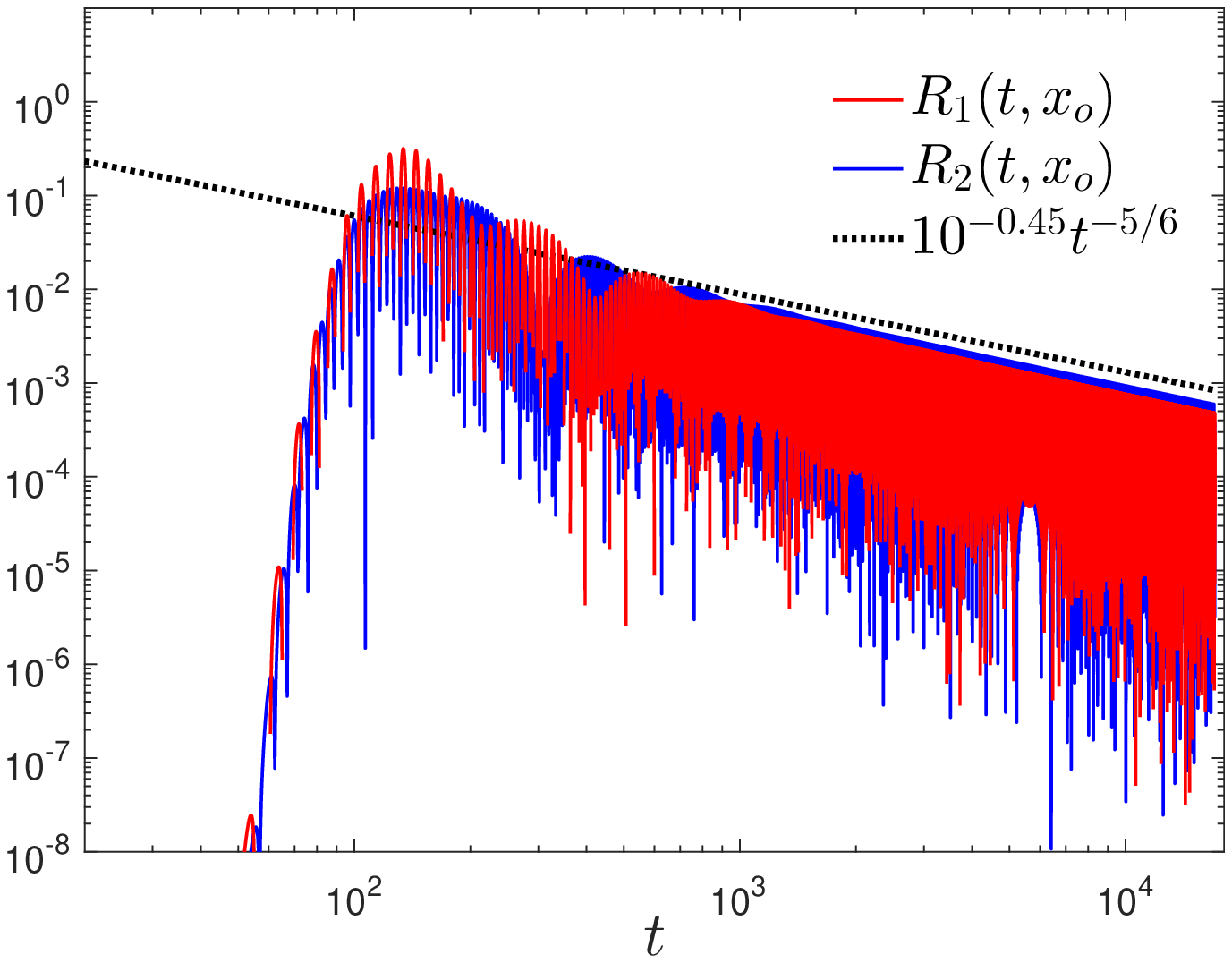}}
	\subfigure{\includegraphics[width=0.4\textwidth]{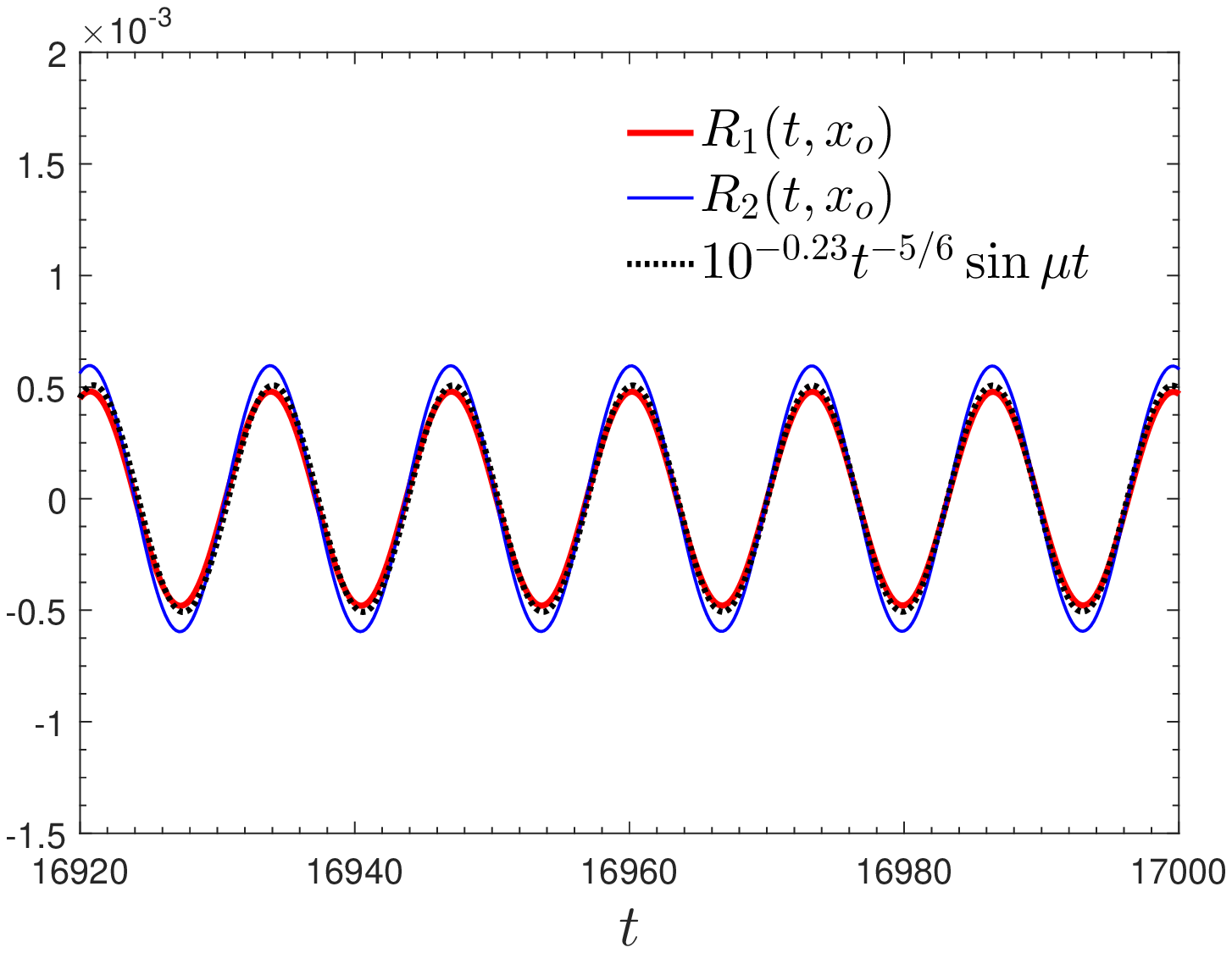}}
	\caption{Left panel: Time evolution of Dirac field on the RN background in the $M\mu<qQ$ regime with $\lambda=1$, $r_-=0.8$ and $\mu=0.48$. The charge of the Dirac field is chosen such that $qQ=1.2M\mu$. The initial wave packet and observation position are the same as those in Fig.\ref{fig: time-domain}. Right panel: Late time behavior of the Dirac field.}
	\label{fig: late-time tail}
\end{figure*}

We remark that parameters of the results in Fig.\ref{fig: time-domain} obey $M\mu>qQ$. In the $M\mu<qQ$ regime, no bound states could be excited in time evolutions. In this case, the Dirac field decays at asymptotically late time as \cite{PhysRevD.72.027501}
\begin{equation}\label{Eq: late-time tail}
	\mathrm{Re}(R_{1,2})\sim t^{-5/6}\sin\mu t,\;\;t\rightarrow\infty.
\end{equation}
This result is confirmed in Fig.\ref{fig: late-time tail}, where we present the time evolution of the Dirac field with $qQ=1.2M\mu$.

\section{Discussion and Conclusion}\label{Sec: conclusion}
We have solved the Dirac equation on the RN background numerically both in frequency and time domain. Our results indicate that quasibound states of massive Dirac fields can only exist in the $M\mu>qQ$ regime, and the long-lived modes exist when $M\mu\gtrsim qQ$, although there is no stationary configuration of Dirac field in this background~\cite{Finster:1998ak}. Here, we shall show that this is actually the necessary condition for the existence of quasibound states. From Eq.(\ref{Eq: effective potential}), at large distance, the effective potential goes as
\begin{equation}\label{Eq: potential V}
V(r)=\mu^2-\frac{2(2M\omega^2-qQ\omega-M\mu^2)}{r}+\mathcal{O}\left(\frac{1}{r^2}\right).
\end{equation}
Notice that this asymptotic behavior is the same as that of the potential of a massive scalar field in the Kerr-Newman spacetime to the first order \cite{PhysRevD.94.064030}. To support quasibound states, the effective potential must have a trapping well outside the BH, and its asymptotic derivative must be positive, i.e., $\frac{dV}{dr}\rightarrow0^+$ as $r\rightarrow\infty$ \cite{Hod:2012zza}. From Eq.(\ref{Eq: potential V}), the necessary condition for a potential well is given by
\begin{equation}\label{Eq: condition for potential well}
\omega>\frac{qQ}{4M}+\sqrt{\frac{\mu^2}{2}+\frac{q^2Q^2}{16M^2}}\equiv f(\mu,q).
\end{equation}
Note also that quasibound states decay exponentially at spatial infinity, which are characterized by $\omega^2<\mu^2$. Thus the necessary condition for the quasibound states is
\begin{equation}\label{Eq: condition for bound states}
f(\mu,q)<\omega<\mu,
\end{equation}
which suggests that $f(\mu,q)$ should be smaller than $\mu$. That is, 
\begin{equation}
	\frac{qQ}{4M}+\sqrt{\frac{\mu^2}{2}+\frac{q^2Q^2}{16M^2}}<\mu.
\end{equation}
Simplifying the above inequality leads to the relation $M\mu>qQ$, which is necessary for the existence of quasibound states. Indeed, all the quasibound states we have found  obey this relation.

The co-rotating modes of massive Dirac fields around a Kerr BH with the frequency obeying the 'superradiance' condition $\omega<m\Omega_H$ could be very long-lived \cite{0264-9381-32-18-184001}. However, we shall show that such long-lived modes do not exist for Dirac fields around RN BHs, since the 'superradiance' condition
\begin{equation}\label{Eq: 'superradiance condition'}
\omega<\omega_c\equiv\frac{qQ}{r_+}
\end{equation}
is incompatible with the quasibound state condition. First, suppose we have a quasibound state in the 'superradiance' regime Eq.(\ref{Eq: 'superradiance condition'}) for Dirac fields on the RN background. Then, from the bound state condition Eq.(\ref{Eq: condition for bound states}), we must have
\begin{equation}
f(\mu,q)<\omega_c.
\end{equation}
Combining this equation and the definition of $f(\mu,q)$ and $\omega_c$ leads to\footnote{In the last step, we have used $Mr_-<r^2_+$.}
\begin{equation}
M^2\mu^2<\frac{Mr_-}{r_+^2}q^2Q^2<q^2Q^2,
\end{equation}
which is incompatible with the quasibound state condition Eq.(\ref{Eq: condition for bound states}). Thus, it is not possible to find quasibound states in the 'superradiance' regime Eq.(\ref{Eq: 'superradiance condition'}) for massive Dirac fields around RN BHs, let alone long-lived modes. It would be interesting to study whether the co-rotating modes, with the frequency obeying a 'superradiance' condition, exist for Dirac fields on the Kerr-Newman background.

It should be pointed out that in this paper we only treat the Dirac field as a test field in the RN background. It would be more interesting to study the back reaction and the gravitational radiation (and possibly electromagnetic radiation) triggered by the Dirac field. Finally, it is plausible that the long-lived configurations we have found here could be generalized to Fermions with higher spin. All of these possibilities deserve further studies in the future.

\begin{acknowledgments}
	This work is supported in part by Science and Technology Commission of Shanghai Municipality under Grant No. 12ZR1421700.
\end{acknowledgments}
\bibliography{Ref}
\end{document}